# Density profiles and pair correlation functions of hard spheres in narrow slits

B. Götzelmann and S. Dietrich

*Fachbereich Physik, Bergische Universität Wuppertal,*

*D–42097 Wuppertal, Federal Republic of Germany*

## Abstract

A hard sphere fluid confined by hard, structureless, and parallel walls is investigated using a certain version of weighted density functional theory. The density profile, the excess coverage, the finite size contribution to the free energy, the solvation force, and the total correlation function are determined as function of the slit width $L$ for various bulk number densities $\rho_b$. In quantitative agreement with rigorous results the present version of density functional theory yields a constant and large but finite number density profile for the limiting case that $L$ is reduced to the diameter of the hard spheres. Within the Derjaguin approximation the results for the slit geometry allows us to obtain the solvation force between two large hard spheres immersed into a fluid of much smaller hard spheres.

61.20.Ne, 68.15.+e, 68.10.Cr

# I. INTRODUCTION

The knowledge of the structural properties of fluids in confined geometries is important both for applied and basic research. For most applications one has to deal with an ensemble of interconnected pores with irregular sizes and geometries. This severely impedes a quantitative comparison between theoretical predictions and actual experimental data. Typically, in this case only general trends and spatially averaged quantities can be tested reliably. Consequently, in such systems many details of theoretical predictions for confined fluids remain unchecked.

Therefore it is highly welcome that substantial experimental progress has been made to prepare model pores which consist of parallel plates whose surfaces are smooth both on atomic and mesoscopic scales and which are immersed into a fluid reservoir. Such well-defined systems are close realizations of corresponding theoretical models and can serve as a suitable testing ground for the behavior of the more complex systems mentioned above. Inter alia, by varying the distance $L$ between the plates, one can study the crossover from a three-dimensional bulk system to a two-dimensional fluid.

From an experimental point of view the structural properties of a fluid confined to this slit geometry can be probed on various levels. First, one can determine global properties such as the mean density in the slit and the excess density compared with a hypothetical bulk system of the same size. Second, ellipsometry and the reflectivity of light, X-rays or neutrons enable one to determine the density profiles normal to the slit surfaces. Third, atomic force microscopes allow one to monitor the solvation force acting on the two plates which reflects the change of free energy of the confined fluid as function of $L$.

More resently, with the advent of powerful neutron and synchrotron sources a fourth component has been added to the spectrum of experimental techniques. The analysis of the *diffuse* scattering of X-rays and neutrons under grazing incidence gives access to the two-point correlation function of the confined fluid. Combined with the knowledge of the one-point correlation function, i.e. the density profile, this provides a deep insight into the structural changes of fluids induced by their confinement including lateral correlations (see, e.g., Refs. [1–4] and references therein).

The purpose of the present contribution is to provide a first step towards guiding such kind of experiments by calculating the two-point correlation function of a *h*ard-*s*phere fluid between two *h*ard *w*alls (HSHW) based on a weighted *d*ensity-*f*unctional *t*heory (DFT). Since this approach requires as a prerequisite the knowledge of the density profiles, we use this opportunity to compare our results for the one-point correlation function with those obtained previously by different techniques for HSHW; furthermore we pay particular attention to the limit of small values of L and to the discussion of the solvation forces.

Our choice for this model system is motivated by its following virtues:

(i) Due to its simplicity it is particularly well suited for comparisons with simulation data. Systems with soft repulsive or long-ranged attractive interaction potentials pose additional difficulties such as their perturbative treatment in analytic approaches and their unavoidable truncations in simulations.

(ii) Within the framework of DFT long-ranged interactions between the fluid particles are typically incorporated by perturbation theory (see (i)), which needs as a prerequisite the results of the corresponding hard sphere reference systems.

(iii) The investigation of the HSHW model is not only an important step for the study of atomic fluids, but it is also an appropriate model for the description of other physical systems. Under favorable conditions certain colloidal particles between glas plates behave like hard spheres in a slit and can be investigated by means of video microscopy [5]. Confined micelles represent another realization of this model system exhibiting substantial technical and biological interest [6–8].

(iv) The HSHW model is the simplest model which allows one to study a nontrivial dimensional crossover from $d = 3$ to $d = 2$. In this model the spheres loose one degree of freedom when the width $L$ of the slit is reduced to the diameter of the hard spheres. This raises the question whether this system is purely two-dimensional and can be characterized as a hard-disc fluid, or whether the presence of the three-dimensional reservoir requires a different description. Since this is important for the interpretation of experiments with very narrow slits, in Sec. II we also introduce the density-functional theory of two-dimensional systems and provide a careful investigation of this limiting case in Sec. III.

The HSHW model has already been investigated by a variety of techniques. For certain values of the chemical potential and of $L$ simulations [8–13] have provided density distributions and values for the solvation force. The same quantities and in addition the total correlation function have been studied in the framework of integral theories such as the Percus-Yevick approximation (PY) [14–17]. Compared with these methods the DFT is computationally less demanding and also enables one to study the free energy of the system. Simulations face the difficulties that they are restricted to a few selected parameter values and that in the grand canonical ensemble fluids confined to narrow slits exhibit strong fluctuations [18]. As far as the integral theories are concerned it is known that they do not capture interesting phase transitions such as wetting phenomena. Since for future work we are interested in them, we implement a specific form of DFT (*w*eighted *d*ensity *a*pproximation (WDA)) which is known to capture them. Thus in a later stage our approach will enable us to build on the present results for the description of the two-point correlation function close to such interfacial phase transitions. For these reasons it is worthwhile to analyze the HSHW model in terms of DFT.

In Sec. II the DFT is introduced and the WDA used here is specified. In Sec. III we investigate the limit $L \to 2\sigma$. A thorough discussion of the density profiles (Sec. IV) and of the correlation functions (Sec. V) follows. Section VI summarizes our main results.

## II. DENSITY FUNCTIONAL THEORY IN d SPATIAL DIMENSIONS

### A. One- and two-point correlation functions

In thermal equilibrium the structural properties of a d-dimensional inhomogeneous fluid consisting of hard generalized spheres follow from the grand canonical partition function

$$\mathcal{Z}_d(\mu, T; [V(\mathbf{R})]) = 1 + \sum_{N=1}^{\infty} \frac{1}{N! \, \Lambda^{dN}} \int d^d R_1 \ldots d^d R_N$$
$$\exp\left(\beta \int d^d R \{[\mu - V(\mathbf{R})]\hat{\rho}_N(\mathbf{R}; \{\mathbf{R}_i\})\} - \beta \Phi(\{\mathbf{R}_i\})\right) \qquad (2.1)$$

as function of the chemical potential $\mu$ and of the temperature $T = 1/(k_B\beta)$; $\{\mathbf{R}_i\} = \{\mathbf{R}_1, \ldots, \mathbf{R}_N\}$. The particles are exposed to an external potential $V(\mathbf{R})$ which includes the confinement due to the container walls and thus limits the spatial integrations here and in the following. For the pair potential

$$\Phi(\{\mathbf{R}_i\}) = \frac{1}{2} \sum_{\substack{i,j=1 \\ i \neq j}}^{N} \psi(|\mathbf{R}_i - \mathbf{R}_j|) \tag{2.2}$$

with

$$\psi(r) = \begin{cases} \infty & , \quad r < \sigma/2 \\ 0 & , \quad r > \sigma/2 \end{cases} \tag{2.3}$$

$\mathcal{Z}_d$ describes hard spheres, discs, and rods of diameter $\sigma$ for $d = 3, 2$, and $1$, respectively. In terms of the number density operator $\hat{\rho}_N(\mathbf{R}; \{\mathbf{R}_i\}) = \sum_{i=1}^{N} \delta(\mathbf{R} - \mathbf{R}_i)$ the equilibrium density profile of the particles in the presence of the external potential $V(\mathbf{R})$ is given by

$$\rho_d(\mathbf{R}) =: \langle \hat{\rho}_N(\mathbf{R}; \{\mathbf{R}_i\}) \rangle = -\frac{1}{\beta} \frac{\delta \ln \mathcal{Z}_d(\mu, T; [V(\mathbf{R})])}{\delta V(\mathbf{R})}. \tag{2.4}$$

The second derivative yields the total correlation function $h(\mathbf{R}_1, \mathbf{R}_2)$ ($\hat{\rho}_N(\mathbf{R}; \{\mathbf{R}_i\}) \equiv \hat{\rho}(\mathbf{R})$):

$$-\frac{1}{\beta} \frac{\delta \rho_d(\mathbf{R}_1)}{\delta V(\mathbf{R}_2)} = <\hat{\rho}(\mathbf{R}_1)\hat{\rho}(\mathbf{R}_2)> - <\hat{\rho}(\mathbf{R}_1)><\hat{\rho}(\mathbf{R}_2)>$$
$$=: h(\mathbf{R}_1, \mathbf{R}_2)\rho_d(\mathbf{R}_1)\rho_d(\mathbf{R}_2) + \rho_d(\mathbf{R}_1)\delta(\mathbf{R}_1 - \mathbf{R}_2). \tag{2.5}$$

Within the framework of the density-functional theory the equilibrium density profile $\rho_d(\mathbf{R})$ minimizes the grand potential functional [19]

$$\Omega_d([\tilde{\rho}_d(\mathbf{R})]; T, \mu; [V(\mathbf{R})]) = F_{ex}^{(d)}([\tilde{\rho}_d(\mathbf{R})]; T) + F_{id}^{(d)}([\tilde{\rho}_d(\mathbf{R})]; T)$$
$$- \int_{\mathbb{R}^d} d^d R \, (\mu - V(\mathbf{R})) \, \tilde{\rho}_d(\mathbf{R}). \tag{2.6}$$

The ideal gas contribution is given analytically by ($\Lambda$ is the thermal de Broglie wave length)

$$F_{id}^{(d)}([\tilde{\rho}_d(\mathbf{R})]; T) = \frac{1}{\beta} \int_{\mathbb{R}^d} d^d R \, \tilde{\rho}_d(\mathbf{R})[\ln(\tilde{\rho}_d(\mathbf{R}) \, \Lambda^d) - 1]. \tag{2.7}$$

The support of the trial function $\tilde{\rho}_d(\mathbf{R})$ is that domain in $\mathbb{R}^d$ where the external potential $V(\mathbf{R})$ differs from infinity; otherwise $\tilde{\rho}_d(\mathbf{R}) = 0$. The excess Helmholtz free-energy functional $F_{ex}^{(d)}([\tilde{\rho}_d(\mathbf{R})]; T)$ is not known exactly and an appropriate approximation has to be chosen (see, c.f., Sec. II.B). Once the density profile $\rho_d(\mathbf{R})$ has been obtained by minimizing Eq. (2.6) the direct correlation function

$$c_d^{(2)}(\mathbf{R}_1, \mathbf{R}_2; [\rho_d(\mathbf{R})]) := -\beta \frac{\delta^2 F_{ex}^{(d)}[\rho_d(\mathbf{R})]}{\delta \rho_d(\mathbf{R}_1) \, \delta \rho_d(\mathbf{R}_2)} \tag{2.8}$$

and via the Ornstein-Zernicke equation

$$h_d(\mathbf{R}_1, \mathbf{R}_2) = c_d^{(2)}(\mathbf{R}_1, \mathbf{R}_2)$$
$$+ \int_{\mathbb{R}^d} d^d R_3\, c_d^{(2)}(\mathbf{R}_1, \mathbf{R}_3)\, \rho_d(\mathbf{R}_3)\, h_d(\mathbf{R}_3, \mathbf{R}_2) \qquad (2.9)$$

the radial distribution function $g_d(\mathbf{R}_1, \mathbf{R}_2) = h_d(\mathbf{R}_1, \mathbf{R}_2) + 1$ are accessible.

We shall compute all quantities for a slit consisting of two parallel structureless hard walls which are described by the external potential

$$V(\mathbf{R}) = \begin{cases} \infty &, \quad z < \sigma,\ z > L - \sigma \\ 0 &, \quad \sigma < z < L - \sigma \end{cases}. \qquad (2.10)$$

Our choice of the origin and the width $L$ are motivated by the comparison with an atomistic model of a slit. In this case the walls consist of two parallel semi-infinite crystals. The nuclei of the atoms forming the top layer of one of these crystals lie in a plane which is located at $z = 0$ for the left wall and at $z = L$ for the right wall. (We do not consider vicinal surfaces.) Between these two walls a fluid with a soft interaction with the substrate has a nonvanishing number density for $0 < z < L$. If the atoms forming the walls are replaced by hard spheres of diameter $\sigma$ and are smeared out in the lateral directions, a fluid with a hard sphere interaction is exposed to the potential defined in Eq. (2.10). Since this substrate potential is translationally invariant with respect to the lateral coordinates $x$ and $y$ (in $d = 3$), the density profile $\rho_3(\mathbf{R})$ depends only on the normal coordinate $z$ as long as there is no freezing which leads to a periodic density variation also in the lateral directions [20]. The total and the direct correlation function depend on the normal distances $z_1$ and $z_2$ from the wall and on the lateral distance $r_{12} = |\mathbf{r}_1 - \mathbf{r}_2|$ where $\mathbf{R} = (\mathbf{r}, z) = (x, y, z)$. (In the case of freezing the two-point correlation functions depend on $\mathbf{r}_{12} = \mathbf{r}_1 - \mathbf{r}_2$ instead of $|\mathbf{r}_1 - \mathbf{r}_2|$.)

### B. The linear weighted density approximation

Since the exact expression for the Helmholtz free-energy functional is not known one has to resort to one of the approximations known in the literature [21]. Depending on the physical system and the quantities under consideration one chooses that approach which captures the essential features without being computationally too demanding.

For a hard sphere fluid ($d = 3$) close to a single hard wall in a previous publication [22] we argued that the calculation of density profiles and of correlation functions can be carried out successfully using the linear weighted density approximation (LWDA) [23]. In this approach four weighted densities

$$\bar{\rho}_\nu(\mathbf{R}_1) = \int_{\mathbb{R}^3} d^3 R_2\, w_\nu(|\mathbf{R}_1 - \mathbf{R}_2|)\, \rho(\mathbf{R}_2) \qquad (2.11)$$

with normalized weights ($\omega = \frac{\pi}{6}\sigma^3$)

$$w_\nu(R) = \frac{1}{8\omega}\Theta(\sigma - R)\begin{cases} 1 &, \quad \nu = 0 \\ (1 + \frac{3}{\nu})(1 - \frac{R^\nu}{\sigma^\nu}) &, \quad \nu = 1, 2, 3 \end{cases} \qquad (2.12)$$

are introduced. The excess free energy is a functional of these weighted densities,

$$F_{ex}^{(d=3)}[\rho] = \sum_{\nu=0}^{3} \int_V d^3R \left\{ f_\nu(\bar{\rho}_\nu(\mathbf{R})) \right.$$
$$\left. + \frac{1}{2}[\rho(\mathbf{R}) - \bar{\rho}_\nu(\mathbf{R})] f'_\nu(\bar{\rho}_\nu(\mathbf{R})) \right\} , \quad (2.13)$$

where the functions

$$\frac{\beta\omega}{\eta} f_\nu(\eta) = \begin{cases} -16 + 4(1 - \frac{4}{\eta})\log(1-\eta) & \nu = 0 \\ \frac{3(-16 + 26\eta - 7\eta^2)}{2(1-\eta)^2} + 3(1 - \frac{8}{\eta})\log(1-\eta) & \nu = 1 \\ 0 & \nu = 2 \\ \frac{40 - 68\eta + 25\eta^2}{(1-\eta)^2} - 8(1 - \frac{5}{\eta})\log(1-\eta) & \nu = 3 \end{cases} \quad (2.14)$$

depend on the dimensionless packing fraction $\eta = \omega \rho$. By construction in the limit of a homogeneous density distribution ($\rho(\mathbf{R}) = \rho_b$) the LWDA free energy $F_{ex}^{(d=3)}[\rho_{d=3}]$ and the corresponding direct correlation function $c_{d=3}^{(2)}(\mathbf{R}_1, \mathbf{R}_2)$ reduce to the known Percus-Yevick (PY) bulk results [24]. This is important as we need a proper bulk limit in order to be able to describe correctly the influence of the walls. Furthermore this allows us to express the results of a slit, whose thermodynamic state is characterized by the intensive variable $\mu$, equivalently in terms of the density $\rho_b$ of a bulk fluid with the same chemical potential $\mu$. This facilitates the comparison with previous publications in which the results for similar geometries are expressed in terms of $\rho_b$ [22]; in addition dependences on $\rho_b$ are easier to interpret than those on $\mu$.

There seem to be only very few WDA which are specifically designed to describe an inhomogeneous hard disc fluid [25]. This dearth is tied to the fact that experimental results are rare, as it is difficult to realize a truly two-dimensional system experimentally and that the construction of many WDAs relies on the knowledge of an analytic expression for the bulk fluid free energy and of the direct correlation function. In three-dimensional systems of spheres the Percus-Yevick closure can be used, but there is no analytic solution thereof known for the two-dimensional case. (For an alternative approach see Ref. [25].) Although it should be possible to construct a WDA also for this case following the concept of Curtin and Ashcroft [26] which does not require *analytic* expressions for the bulk quantities, it is natural to analyze an alternative approach [27] which amounts to evaluate the excess free-energy functional $F_{ex}^{(d=3)}[\rho_{d=3}]$ (Eq. (2.13)) of the hard sphere fluid for

$$\rho_3(\mathbf{R}) = \rho_2(x,y)\,\delta(z) \quad (2.15)$$

and which leads to the following approximate expression for the two-dimensional hard disc fluid in an area $A$:

$$F_{ex}^{(2)}[\rho_2] = \sum_{i=0}^{3} \int_A dx\,dy \left\{ \frac{1}{2}\rho_2(x,y) f'_\nu(\bar{\rho}_\nu(x,y,z=0)) \right.$$
$$\left. + \int_{-\sigma}^{\sigma} dz \left[ f_\nu(\bar{\rho}_\nu(\mathbf{R})) - \frac{1}{2}\bar{\rho}_\nu(\mathbf{R}) f'_\nu(\bar{\rho}_\nu(\mathbf{R})) \right] \right\}, \text{ LWDA}. \quad (2.16)$$

The fact that this approximation originates from a theory designed for three spatial dimensions is especially apparent in the weighted densities

$$\bar{\rho}_\nu(\mathbf{R}) = \int_{\mathbb{R}^2} dx' dy' \rho_2(x', y') \, w_\nu(\sqrt{|x-x'|^2 + |y-y'|^2 + z^2}), \tag{2.17}$$

which still depend on three coordinates.

## III. COMPARISON OF A HARD DISC FLUID WITH A HARD SPHERE FLUID IN A NARROW SLIT

In order to assess the quality of the approximation leading to the free energy functional in Eq. (2.16) of a hard disc fluid we consider the special case of a homogeneous density distribution. By setting $\rho_2(x,y) = \rho_{2,b}$ in Eq. (2.15) the excess free energy in Eq. (2.16) can be compared with the results of the scaled-particle theory (SPT) [28]

$$\beta F_{ex}^{(2)}[\rho_{2,b}] = A \rho_{2,b} \left[ \frac{\eta_2}{1-\eta_2} - \ln(1-\eta_2) \right], \text{ SPT}, \tag{3.1}$$

where $A$ denotes the cross section of the slit and $\eta_2 = \rho_{2,b} \frac{\pi}{4}\sigma^2$ the packing fraction. It turns out that for all densities the values of the LWDA free energy (Eq. (2.16)) is higher than the SPT results and the difference increases with increasing density $\rho_{2,b}$. For $\rho_{2,b}\sigma^3 = 0.6$ there is a deviation of about 17%. The pressure of the system is given by

$$P = -\left.\frac{\partial F^{(2)}[\rho_2]}{\partial A}\right|_{T,N} = \rho_d^2 \left.\frac{\partial (F^{(2)}[\rho_2]/N)}{\partial \rho_2}\right|_{T,N} \tag{3.2}$$

and can be compared with results of integral theories such as the hypernetted chain approximation (HNC) and simulations [29]. The pressure calculated within the LWDA approximation is comparable to that of the HNC results but higher than the one obtained from simulation data. Thus we conclude that the suggested functional in Eq. (2.16) is a reasonable but not very accurate approximation for a hard disc fluid.

If the width of a slit filled with a hard sphere fluid is decreased one could be inclined to expect that in the limit $L \to 2\sigma$ (compare Eq. (2.10)) the density profile reduces to a $\delta$-distribution as in Eq. (2.15) and consequently that in this limit the system is described by the density-functional theory of a two-dimensional system, e.g. by the one proposed above. However, if Eq. (2.15) is inserted into the expression for the ideal gas contribution to the free energy functional (Eq. (2.7) for $d = 3$), one obtains a mathematically ill-defined expression. Since this defect is not cured by the LWDA excess free-energy functional, a well-defined grand canonical functional $\Omega[\rho]$ (Eq. (2.6)) can only be constructed, if $\rho(\mathbf{R})$ remains a finite function even in the limit $L \to 2\sigma$. Indeed Henderson [30] has shown that in first order of $\tilde{L} := L - 2\sigma$ the contact density is given exactly by

$$\rho(z = \sigma^+) = \Lambda^{-3} \exp(\frac{\mu}{kT})[1 - \rho(\sigma)\pi\sigma^2\tilde{L} + O\left(\left(\frac{\tilde{L}}{\sigma}\right)^2\right)] \tag{3.3a}$$

$$= \Lambda^{-3} \exp(\beta\mu) \frac{1}{1 + \pi\sigma^2\tilde{L}\Lambda^{-3}\exp(\beta\mu)} + O\left(\left(\frac{\tilde{L}}{\sigma}\right)^2\right). \tag{3.3b}$$

This implies that in the grand canonical ensemble in the limit $L \to 2\sigma$ the fluid is squeezed out of the slit and that the number density $<N>/A$ of the particles per area still contained inside the slit vanishes linearly as the width is decreased: $<N>/A = \int_\sigma^{L-\sigma} dz \rho(z) \overset{\tilde{L}\to 0}{\to} \rho(z=\sigma^+)\tilde{L}$. Due to this small number of particles per area the fluid behaves like an ideal gas and in zeroth order the density is determined by the Boltzmann distribution ( Eq. (3.3b)). On the other hand the local number density is rather high because the value of the chemical potential $\mu$ is imposed by the bulk reservoir. It is interesting to note that one obtains the same limit for rods confined to a finite line segment in the limit of a vanishing length of the segment [31]: $<N>/\tilde{L} \overset{\tilde{L}\to 0}{\to} \frac{1}{\Lambda}\exp(\mu\beta)$.

In order to investigate the limit $\tilde{L} \to 0$ within the framework of the LWDA the functionals in Eq. (2.13) and in Eq. (2.7) are simplified according to the following approximations. For small values of $\tilde{L}$ the local density in the slit can be taken to be constant and equal to $\rho(\sigma)$. Also the weights $w_\nu(\sqrt{|x_1-x_2|^2 + |y_1-y_2|^2 + |z_1-z_2|^2})$ do not vary significantly for $\sigma < z_1 < L-\sigma$ for $\nu = 0, 1, 2, 3$ and $x_1, x_2, y_1, y_2, z_2 \in \mathbb{R}$ fixed. With the resulting functionals

$$F^{(3)}_{ex}[\rho(\sigma)] = \sum_{i=0}^{3} \int_A dx\,dy \left\{ \frac{1}{2}\rho(\sigma)\tilde{L}f'_\nu(\bar{\rho}_\nu(x,y,z=0)) \right.$$
$$\left. + \int_0^L dz \left[ f_\nu(\bar{\rho}_\nu(\mathbf{R})) - \frac{1}{2}\bar{\rho}_\nu(\mathbf{R})f'_\nu(\bar{\rho}_\nu(\mathbf{R})) \right] \right\}, \quad (3.4)$$

$$\bar{\rho}_\nu(\mathbf{R}) = \int_{\mathbb{R}^2} dx'dy'\rho(\sigma)\tilde{L}\,w_\nu(\sqrt{|x-x'|^2 + |y-y'|^2 + z^2}), \quad (3.5)$$

and

$$\beta F^{(3)}_{id}[\rho(\sigma)] = A\,\tilde{L}\rho(\sigma)\left[\ln(\Lambda^3\rho(\sigma)) - 1\right] \quad (3.6)$$

the grand canonical potential

$$\beta\,\Omega[\rho(\sigma)] = \beta\left\{F^{(3)}_{ex}[\rho(\sigma)] + F^{(3)}_{id}[\rho(\sigma)]\right\} - \beta\mu A\tilde{L}\rho(\sigma) \quad (3.7)$$

is determined. Since $\rho(\sigma)$ remains finite, $\tilde{L}\rho(\sigma)$ vanishes in the limit $\tilde{L} \to 0$ and the functions $f$ in Eq. (3.4) can be expanded into Taylor series yielding

$$\frac{\beta}{A}\Omega[\rho(\sigma)] = \frac{1}{2}\sigma^2\pi\rho(\sigma)^2\tilde{L}^2 + \rho(\sigma)\tilde{L}\left[\ln(\Lambda^3\rho(\sigma)) - 1\right] - \mu\tilde{L}\rho(\sigma)\beta,\ \tilde{L}\to 0. \quad (3.8)$$

Minimizing this expression with respect to the contact density $\rho(\sigma)$ leads to

$$\Lambda^3\rho(\sigma) = \exp(\mu\beta - \rho(\sigma)\,\pi\sigma^2\tilde{L}) \quad (3.9)$$

and via further expansion to Eq. (3.3a). This renders the satisfying result that in the limit $\tilde{L} \to 0$ up to first order in $\tilde{L}$ the LWDA reduces to the exact result.

Since as shown above the LWDA is capable of both describing reasonable well a hard disc fluid and reproducing correctly in first order of $\tilde{L}$ the limit $\tilde{L} \to 0$, for a hard sphere fluid we are in the position to compare these two different physical systems within one and the same approach. The excess free energy functionals in Eq. (2.16) and in Eq. (3.4) can be mapped onto each another using the replacement $\rho_{2b} \leftrightarrow \tilde{L}\rho(\sigma)$. This replacement only states, that the number densities per area have to be the same for both systems:

$$\int_{\sigma}^{\sigma+\tilde{L}} dz\, \rho(z) = \tilde{L}\,\rho(\sigma) + O(\tilde{L}^2) = \rho_{2,b}. \tag{3.10}$$

Obviously this mapping cannot be used to relate the ideal gas contributions of a hard disc fluid (Eq. (2.7) with d=2) and of a fluid in a slit (Eq. (3.6)). For them no simple map is found. Thus we conclude that the hard sphere fluid within narrow slits and connected to a reservoir does not resemble the genuinely two-dimensional hard disc fluid. If one wants to prepare a quasi two-dimensional system with a non-vanishing density $\rho_{d=2}$ one has to resort to the canonical ensemble, i.e., one has to restrict a fixed number of particles to a finite volume $V = A\sigma$.

The mechanism, which leads to a finite density in the limit $\tilde{L} \to 0$, is revealed by Eq. (3.8). The grand canonical potential is split into the excess part $\frac{1}{2}\sigma^2 \pi \rho(\sigma)^2 \tilde{L}^2$ which captures the contribution of the interaction between the particles, the ideal gas part $\rho(\sigma)\tilde{L}[\ln(\Lambda^3 \rho(\sigma))-1]$ which mainly takes into account the entropy, and the part $-\mu \tilde{L}\rho(\sigma)\beta$ due to the chemical potential. As the excess part is quadratic in $\tilde{L}$ it becomes less important as the width of the slit decreases and ultimately the ideal gas part determines the behavior of the system. Thus the entropy is responsible for the fact that the local density in the slit remains finite in limit $\tilde{L} \to 0$.

## IV. CHARACTERIZATION OF HARD SPHERE FLUIDS IN NARROW SLITS

Since lateral ordering phenomena are beyond the scope of the present work, we focus on sufficiently small densities which are below the onset of such freezing transitions. In the homogeneous bulk freezing occurs at $\rho_b \sigma^3 \simeq 0.94$ [32], but already near a single wall prefreezing sets in at a slightly lower density [20]. Experiments in slits [33,34] revealed a rich structure including phase transitions between different ordered states as the width of the slit varies. For these kind of systems canonical Monte Carlo simulation were able to reproduce the phase diagram satisfactorily [35]. By analyzing the total correlation function of Monte Carlo simulation data Chu et al. [8] have shown, that for $L = 3\sigma$ a hexagonal packing close to the walls occurs. Thus beside a complete freezing of the whole slit also lateral ordering in parts of the slit close to the walls seems possible. Therefore we have limited our present investigations to $\rho_b \sigma^3 = 0.68$. For this density extensive Monte Carlo simulations [11] gave no hints for an onset of lateral ordering.

### A. Properties of the density profile

Using various mesh sizes $(0.005\sigma \ldots 0.05\sigma)$ for the integration the grand potential in Eq. (2.6) was minimized for numerous slit widths ($L = 2.001\sigma \ldots 12\sigma$). If the width $L$ is larger than $12\sigma$, the resulting density profiles close to one of the walls agree well with those near the single wall of the corresponding semi-infinite system [22]. In this limit the wall theorem ($\rho(\sigma^+) = \beta P$) is fullfilled. Figure 1 shows the density profile for two different slit widths ($L = 5.1\sigma$ and $L = 3.8\sigma$). They are symmetric and exhibit a layered structure due to packing effects. Due to the absence of rigorous results the accuracy of our results can only be assessed by comparing them with simulation data. Surprisingly to our knowledge there is only a single publication, namely a molecular dynamics simulation (MD) [13], which allows us a *quantitative* comparison. In this case the chemical potential of the fluid is not known and one has to use a different parameter in order to be able to map the two approaches onto each another. Since the LWDA is not an exact theory, the degree of agreement depends on the choice of this parameter. Here we use the mean density $\rho_m$ (see, c.f., Eq. (4.3)) of the particles in the slit, but we correct the value of the LWDA by a factor $\rho(z=\sigma^+)_{simulation}/\rho(z=\sigma^+)_{LWDA} = \frac{P_{CS}}{P_{PY}} = 1.03$ in order to take into account the fact that the LWDA slightly underestimates the pressure of the bulk fluid. $P_{PY}$ denotes the pressure of a homogeneous LWDA fluid and $P_{CS}$ is the almost exact Carnahan-Starling pressure [36]. Figure 1(a) reveals a satisfactory agreement with the simulation data. Similar to the case of a hard sphere fluid close to a single wall of a semi-infinite system [22], the first minimum is slightly too shallow, but the phases of the oscillations agree rather well. If one investigates the changes in the form of the density profile upon increasing the slit width $L$, one finds the following scenario: At very small widths the contact density $\rho(\sigma)$ is very high and the profile between the walls is almost flat. If $L$ is increased the profile develops a U-shape with a single minimum at $z = L/2$ and both $\rho(\sigma)$ and $\rho(L/2)$, which are shown in Fig. 2 as the full and the dotted line, respectively, decrease rapidly. As function of $L$ the contact value attains a minimum at $L = 2.39\sigma$ and reaches again a maximum at $L = 3.01\sigma$. As function of $L$ the value of the density $\rho(L/2)$ in the center reaches its first minimum at $L = 3.07\sigma$ and then increases slowly. At about $L = 3.4\sigma$ two local maxima in the density profile $\rho(z)$ appear at approximately $z = 2\sigma$ and $z = L - 2\sigma$ (compare Fig. 1(b)) which are each approximately one hard sphere diameter $\sigma$ apart from the first layer at the distant wall. These two extrema merge into a single maximum when the slit width is further increased ($L = 4.0\sigma$). It is difficult to determine reliably the precise value of the slit width for which these two extrema can no longer be distinguished because it depends sensitively on the approximations entering LWDA [37]. For $L = 3.98\sigma$ the contact density attains its third maximum and the density profile has a pronounced W-like shape. For increasing $L$ the peak in the center broadens and starting at about $L = 4.5\sigma$ it splits into two peaks located approximately at $z = 2\sigma$ and $z = L - 2\sigma$. In between there is a local minimum, which deepens and is smallest for $L = 5.1\sigma$ (see Fig. 2). The slit contains now four layers. If the width is further increased more layers are added by a similar mechanism. The extrema characterizing this process are given in Table I. The values of the contact density and of various other quantities which will be defined below (see Fig. 3) oscillate as a function of $L$ with a period of about $\sigma$. Within each of these oscillations another layer is added to the slit. The density $\rho_L(L/2)$ at the center of the slit has a minimum if an even number of layers are in the slit and a maximum if there is an odd number. Therefore this quantity exhibits a periodicity of about $2\sigma$.

A useful global description of the density distribution in the slit is given by the excess coverage

$$\Gamma(L) := \int_0^L dz\, [\rho(z) - \rho_b], \tag{4.1}$$

and the related mean density

$$\rho_m(L) := \frac{1}{L} \int_0^L dz\, \rho(z), \tag{4.2}$$

or

$$\rho_m(L) = \rho_b + \Gamma(L)/L. \tag{4.3}$$

As shown in Fig. 3 $\Gamma(L)$ exhibits oscillations with a period of about $\sigma$. Due to the factor $1/L$ (see Eq. (4.3)) the corresponding oscillations of $\rho_m(L)$ are less pronounced.

In the limit $L \to \infty$ the coverage $\Gamma(L)$ equals twice the coverage of a hard sphere fluid close to a single wall which has been discussed in Ref. [22]. For $L$ larger than $5\sigma$ the coverage differs only slightly from the limiting value $\Gamma(\infty)$:

$$\frac{\Gamma(L) - \Gamma(\infty)}{\Gamma(\infty)} < 0.04, \quad L > 5\sigma. \tag{4.4}$$

This specifies the range of validity for the well known approximation $\rho_m \simeq \rho_b + \Gamma(\infty)/L$ [38].

In the limit $L \to 2\sigma$ the density $\rho(z)$ is constant (Eq. (3.3b)) so that Eq. (4.2) yields

$$\Gamma(L \to 2\sigma)) = [L - 2\sigma]\left[\Lambda^{-3} \exp(\beta\mu) - \rho_b\right] + O\left(\left(\frac{\tilde{L}}{\sigma}\right)^2\right) \stackrel{L \to 2\sigma}{\to} 0. \tag{4.5}$$

### B. Finite size contribution to the free energy and resulting solvation forces

In the context of a slit geometry a particularly interesting quantity is the finite size contribution to the grand potential,

$$\gamma(L) := \lim_{A \to \infty} \frac{1}{A} \left(\Omega[\rho] + [L - 2\sigma] P_{PY}\right), \tag{4.6}$$

where $P_{PY}$ is the pressure of a homogeneous bulk liquid at the same chemical potential $\mu$. In the limit $L \to \infty$ $\gamma(L)$ reduces to twice the surface tension of a hard sphere fluid close to the single hard wall of a semi-infinite system. Inserting Eq. (3.9) into the expansions in Eqs. (3.8) and (4.6) yields in the opposite limit $L \to 2\sigma$

$$-\gamma(L) = \left[\frac{1}{\beta}\rho(\sigma) - P_{PY}\right][L - 2\sigma] + O\left(\left(\frac{\tilde{L}}{\sigma}\right)^2\right). \tag{4.7}$$

The behavior of $\gamma(L)$ in the intermediate regime between these two limits is displayed in Fig. 3. The difference $\gamma(L) - \gamma(\infty)$ decreases oscillatorily with increasing slit width. The maxima decay exponentially ($\sim \exp(-1.23\tilde{L}/\sigma)$); their positions are given in Tab. I.

The force between the two plates is an experimentally accessible quantity [6]. Based on thermodynamics this so-called solvation force per area $f(L)$ is given as [39]

$$f(L) = -\lim_{A \to \infty} \frac{1}{A} \left(\frac{\partial \Omega}{\partial L}\right)_{T,\mu,A} - P_{PY}. \tag{4.8}$$

Using Eq. (4.6) one obtains

$$f(L) = -\left(\frac{\partial \gamma}{\partial L}\right)_{T,\mu,A} \tag{4.9}$$

so that

$$\gamma(L) - \gamma(\infty) = -\int_\infty^L dL' \, f(L'). \tag{4.10}$$

Since $\gamma(L \to 2\sigma) = 0$ (Eq. (4.7)) it follows that

$$\gamma(\infty) = -\int_{2\sigma}^\infty dL' \, f(L') = 2\gamma_{sl}. \tag{4.11}$$

Thus the surface tension $\gamma_{sl} = \gamma(\infty)/2$ of a hard sphere fluid close to a hard wall is a measurable quantity accessible to force measurements. Equation (4.11) remains valid even for softly interacting spheres close to a hard wall, but not if the hard wall is replaced by a soft substrate potential.

The solvation force can also be expressed in terms of the difference between the contact density at the finite slit width $L$, $\rho(\sigma)$, and at infinite slit width, $\rho_\infty(\sigma)$, (see Ref. [40] and the Appendix):

$$\beta f(L) = \rho(\sigma) - \rho_\infty(\sigma). \tag{4.12}$$

This difference is shown in Fig. 2. Minimizing the grand canonical potential within LWDA we obtain both the density profile as the minimizing function, which leads to the force via Eq. (4.12), and the value of the minimum $\Omega[\rho]$ which yields the force using Eq. (4.8). We find numerically that both routes lead to the same result. This can be anticipated because any WDA is thermodynamically self-consistent with respect to this relation (see the Appendix). Here it serves as a helpful check of the numerical calculations.

For a physical understanding of the above results it is rewarding to consider the system depicted in Fig. 4. The solvation force per area $f(L)$ is the net force exerted on the wall $b$ and positive if it is directed outwards, i.e. to the right. For $L > 2\sigma$ it is shown in Fig. 2 and for $\sigma < L < 2\sigma$ it is constant,

$$f(L) = -\frac{1}{\beta} \rho_\infty(\sigma), \qquad \sigma < L < 2\sigma, \tag{4.13}$$

because the particles on the right side of wall $b$ exert the constant bulk pressure $P_{PY} = \frac{1}{\beta}\rho_\infty(\sigma)$ to the left. Starting with the wall $b$ at a position corresponding to a slit width $L$ the work $\gamma(L) - \gamma(\infty)$ is gained (Eq. (4.10)) if one moves it to infinity (see Fig. 3). If one now considers the particular case $L = 2\sigma^-$ the above process starts from a configuration involving only a *single* surface and leads to a configuration of three independent solid-fluid interfaces without changing the *bulk* contribution to the free energy of the total system but increasing the *surface* contribution to the free energy by $2\gamma_{sl}$. This provides a transparent interpretation of Eq. (4.11) because the integral over the solvation force is the work applied to the system during this process. In addition these considerations show that Eq. (4.11) is valid in general and not only within LWDA. Together with the general relation in Eq. (4.10) this implies that the equation $\gamma(L \to 2\sigma) = 0$ is also valid in general. Finally, since $\gamma_{sl}$ is negative [22] these arguments also tell that one *gains* work by moving the wall b from $L = 2\sigma$ to $L = \infty$. Thus the generation of these two additional solid-fluid interfaces is favorable.

The negative sign of $\gamma_{sl}$ may provoke the question whether the hard sphere fluid is actually stable against the spontaneous formation of cavities. The effect of the formation of such cavities on the density distribution of the fluid can be thought of to be same as the effect induced by the immersion of, e.g., a hard wall (Fig. 4) with a minimum thickness $\sigma$. The difference in free energy $\Delta F$ between the homogeneous and the corresponding perturbed system consists of the two surface contributions $\gamma_{sl}A$ and the bulk free energy density times the excluded volume $\sigma A$ of the cavity: $\Delta F/A = 2\gamma_{sl} + P_{PY}\sigma$. Although $\gamma_{sl}$ is negative it turns out that $\Delta F$ is positive [41] so that the cavity formation is disfavored.

The above discussion is concerned with the particular cases $L = 2\sigma$ and $L = \infty$. For general $L$ it is worthwhile to note that the extrema of $\gamma(L)$ correspond to the zeros of the solvation force $F(L)$ which are documented in Table I. If in Fig. 4 the wall b is allowed to float freely, the minima of $\gamma(L)$ correspond to the most favorable slit widths. According to Fig. 3 the global minimum is located at $L = 2.18\sigma$. Thus in thermal equilibrium the optimum configuration in Fig. 4 corresponds to the case in which the walls a and b are separated such that a monolayer of hard spheres fits in between with a little bit space left. However, one should keep in mind that this statement is only valid if the mass of the piston is much larger than the masses of the hard spheres. Otherwise the fluctuations of the position of the piston must be treated on the same footing and together with those of the hard spheres. In this sense the above line of argument, i.e. that the equilibrium position of the piston is determined by the minimum of $\gamma(L)$, corresponds to a Born-Oppenheimer approximation.

At the width $L = n\sigma$ just $n - 1$ spheres fit on top of each other into the slit. One may wonder whether this peculiar matching condition leaves a particular signature in the $L$-dependence of the various quantities studies above. The corresponding $(d-2)$-dimensional problem of hard rods of length $\sigma$ confined to a segment of length L on a line can be solved exactly [31] and one finds in this case that the second derivative of the mean number of rods exhibits discontinuities at $L = n\sigma$ whose magnitudes decrease for increasing values of L [31]. Since the additional spatial dimensions of a three-dimensional slit allow for an easier rearrangement of the spheres upon packing, we expect that in $d = 3$ these discontinuities are either smeared out or shifted to higher derivatives. Although in principle one should be prepared for the occurrence of such singularities in, e.g., $\gamma(L)$ or $f(L)$ they turn out to be so weak that they do not show up in our present LWDA approach on the scale of the numerical resolution we used.

### C. Derjaguin approximation for the force between large spheres

So far our analysis has been confined to the study of the slit geometry which may be applicable to force microscope measurements of confined colloidal particles whereby the solute particles only contribute to the effective interaction between them; in the present context this effective interaction is approximated by a hard-core repulsion potential. However, it turns out that the results for the slit geometry can be even used to analyze this latter effective interaction potential between large colloidal particles of radius $R$ which are immersed into a solute composed of small particles with diameter $\sigma$ [42]. (For $R \to \infty$ this problem reduces to the standard slit geometry.) If the centers of the two large hard spheres are kept at a fixed distance $2R + h - \sigma$, in the limit $R \gg \sigma$ the solvation force between them is given by the Derjaguin approximation [43]:

$$\begin{aligned} f_s(h) &= \pi R \int_h^\infty dL'\, f(L') \\ &= \pi R[\gamma(h) - \gamma(\infty)], \end{aligned} \quad (4.14)$$

where Eq. (4.11) has been used. The large spheres touch each other for $h = \sigma$; in this case there is no small sphere between the two large spheres along the symmetry axis. According to Eq. (4.14) the finite size contribution to the *free energy* of a slit of width $h$ is proportional to the *force* between two hard spheres of radius $R$. For slit widths larger than $2\sigma$ the free energy is given by Fig. 3 and for $\sigma < L < 2\sigma$ Eq. (4.13) is used in Eq. (4.14). The combination of these results leads to Fig. 5.

The global minimum at $h = \sigma$ indicates that a strong depletion force will press the two spheres together if they touch each another. However, in order to find the thermodynamically most favorable separation one must consider the effective interaction potential

$$W_s = \int_\infty^h dh'\, f_s(h') \quad (4.15)$$

between the two large spheres. This is shown in Fig. 5 as the dotted line. For low densities ($\rho\sigma^3 < 0.2$) this quantity has been investigated in the framework of an expansion into powers of $\rho$ [44]. The present density-functional theory extends these results to higher densities. Inter alia, the interest in this effective potential arises from the question whether a binary mixture of hard spheres can exhibit flocculation. For such systems the PY theory rules out phase separation at all densities and size ratios [45]. However, modern integral theories indicate that phase separation can occur [46].

## V. TWO-POINT CORRELATION FUNCTION

The two-point correlation function of the HSHW model depends on the normal distances $z_1$ and $z_2$ from the left wall and on the lateral distance $r_{12}$ (see Subsec. II.A). The system is specified by the bulk density $\rho_b$ of the corresponding homogeneous system with the same chemical potential and by the width of the slit $L$. Thus the correlation function depends on five independent variables. Within the context of a research paper a complete graphical account of the dependences on all five variables is not feasible. Therefore we have decided to discuss the general mechanism governing the behavior of the total correlation function on the basis of the Percus test-particle theorem (see the following paragraph) and to select the display of the dependence on $r_{12}$ (with $z_1 = z_2$ and fixed $\rho_b$ in all plots) for $L$ fixed and various values of $z_1$ as well as for $z_1$ fixed and various values of $L$. Here and in the following we discuss the case $d = 3$ only.

In the present context the Percus test-particle theorem [40] states that the product $\rho(z_1)g(r_{12}, z_1, z_2)$ (see Eq. (2.9)) for a hard sphere fluid in a slit of width $L$ is equal to the one-point conditional density distribution $\rho(\mathbf{R}_1 = (\mathbf{r}_{12}, z_1)|\mathbf{R}_2 = (\mathbf{0}, z_2))$ of a hard sphere fluid exposed to an external potential consisting of a slit of width $L$ and in addition of a hard sphere of diameter $\sigma$ whose center is fixed at $\mathbf{R}_2 = (\mathbf{0}, z_2)$. For $\rho_b\sigma^3 = 0.546$, $L = 5.1$, and $z_2 = \sigma$ this product is shown in Fig. 6. In the limit $r_{12} \to \infty$ it reduces to the density profile in Fig. 1(a) which corresponds to the same slit width and the same chemical potential. This comparison reveals that approximately $\rho(\mathbf{R}_1 = (\mathbf{r}_{12}, z_1)|\mathbf{R}_2 = (\mathbf{0}, z_2))$ is the superposition of the density profile of Fig. 1(a) and the oscillatory density distribution around a fixed hard sphere placed in a previously homogeneous bulk fluid. Accordingly the coordinates $z_1$ of the maxima and minima of $\rho(\mathbf{R}_1|\mathbf{R}_2)$ which are denoted in Fig. 9 by the dots and circles, respectively, almost coincide with those of the density profile of Fig. 1(a) and thus in Fig. 9 they line up nearly parallel to the wall. This general mechanism was also born out in previous analyses (PY-approximation for a fluid in a slit [16]; LWDA for a fluid close to a single wall [22]) and has proven to yield a roughly correct picture of the radial correlation function in confined geometries.

It is rewarding to investigate for different slit widths $L$ the radial distribution function $g(r_{12}, z_1, z_2)$ as function of $r_{12}$, i.e. parallel to the wall with $z_1 = z_2$ fixed. This reveals the influence of the distant wall on the correlation function close to the near wall. This dependence is of particular interest because it can be measured directly by using digital video microscopy [5]. Although such experimental data are not yet accurate enough to facilitate a quantitative comparison with theoretical results, we surmise that in the near future the experiments will be improved sufficiently. For a bulk density of $\rho_b\sigma^3 = 0.683$ the radial distribution function is shown for $z_1 = z_2 = \sigma$ and $z_1 = z_2 = 1.5\sigma$ in Figs. 7(a) and Fig. 7(b), respectively. Compared with the corresponding radial distribution function of a homogeneous bulk fluid with the same chemical potential (see the dashed–double-dotted lines in Fig. 7) the amplitude of the oscillations is reduced for $z_1 = z_2 = \sigma$ but enhanced for $z_1 = z_2 = 1.5\sigma$. Since already the increase of $z_1$ by the radius of a sphere alters the amplitudes so profoundly, we conclude that accurate measurements of parallel correlations require a spatial resolution in z-direction of about $0.1\sigma$ or better. For increasing $L$ the radial distribution function reduces rapidly to that of the corresponding semi-infinite system [22]. In Figs. 7(a) already for a slit width of $L = 5\sigma$ ($\tilde{L} = 3\sigma$) for all values of $r_{12}$ the influence of the distant wall is no longer visible within the resolution of the plot. This is remarkable because for this width only 4 spheres fit side by side into the slit.

In Sec. IV we have discussed the HSHW model for two sets of the parameters $L$ and $\rho_b$ in terms of the one-point correlation function and related quantities. In Fig. 8 and 9 the same cases are now investigated in terms of the two-point correlation functions as function of $r_{12}$ for various values of $z_1 = z_2$ and for $\rho_b\sigma^3 = 0.546$ and $L/\sigma = 5.1$ in Fig. 8 (compare Fig. 1(a)) and for $\rho_b\sigma^3 = 0.683$ and $L/\sigma = 3.8$ in Fig. 9 (compare Fig. 1(b)). The various values of $z_1$ are chosen to coincide with the positions of the extrema in the corresponding density profile (see Fig. 1). For both systems in the case $z_1 = \sigma$ the contact value $g(r_{12} = \sigma, z_1, z_2 = z_1)$ is strongly reduced as compared to the bulk value $g_{PY}(\sigma)$, whereas for larger distances $z_1$ these contact values are rather close to each other and to the bulk value. This is in accordance with the results obtained for a semi-infinite hard sphere fluid near a single wall [22]. A further analysis of these results reveals that for those values of $z_1$ which correspond to the minima of the density profile these amplitudes of the oscillations in $g(r_{12}, z_1, z_2 = z_1)$ are enhanced whereas for those values of $z_1$ corresponding to the maxima of $\rho(z)$ these amplitudes are reduced. This is surprising because the amplitudes of the oscillations in the bulk correlation function $g_{PY}(r; \rho_b)$ decrease with decreasing $\rho_b$. Thus the natural attempt to approximate the radial distribution function such that in the case $z_1 = z_2$ it reduces to $g(r_{12}, z_1, z_2 = z_1) \approx g_{PY}(r = r_{12}; \rho_b = \rho(z_1))$ is unsuccessful because this approximation renders the opposite tendency of the actual oscillatory behavior at least for the values of the chemical potentials and slit widths considered here. On the basis of a known expression for the bulk correlation function $g(r)$ there have been efforts [47,48] to construct a suitable ansatz for the radial distribution function for a hard sphere fluid close to a hard wall. Since such an ansatz does not incorporate the peculiar behavior of the actual radial distribution function described above, the reliability of these approximations is rather limited.

## VI. SUMMARY

We have obtained the following main results for a fluid of hard spheres of diameter $\sigma$ confined to parallel and structureless hard walls at a distance $L$:

(1) On the basis of the linear weighted density approximation (LWDA), which describes an inhomogeneous three-dimensional fluid in the grand canonical ensemble, we have constructed an approximation for the two-dimensional analogon of a homogeneous hard disc fluid. This approximation compares favorable with simulation data.

(2) As proven by Henderson [30] in the limit that the width $L$ of the slit is reduced such that it can accommodate at most a monolayer of the fluid ($L \to 2\sigma$) the density profile approaches a large but finite constant value. This implies that in this limit the fluid is squeezed out of the slit. Up to first order in $\tilde{L} = L - 2\sigma$ we find that the LWDA reproduces this limit exactly.

(3) The two-dimensional hard disc fluid and the three-dimensional hard sphere fluid confined to a narrow slit have been compared in the grand canonical ensemble. For finite chemical potentials the confined hard sphere fluid does not resemble the genuinely two-dimensional hard disc fluid.

(4) The density profiles $\rho(z)$ calculated within the LWDA compare satisfactorily with simulation data (Fig. 1). The dependence of the contact density $\rho(z = \sigma)$ on the slit width $L$ is close to that obtained from simulation data, although the amplitude of the oscillations in this dependence as obtained from LWDA is slightly smaller than that obtained from the simulations (Fig. 2).

(5) Both within the framework of exact thermodynamics and within LWDA the finite size contribution to the free energy $\gamma(L)$ (see Eq. (4.6) and Fig. 3) represents the potential of the solvation force $f(L)$. The minima of $\gamma(L)$ (see Table I) correspond to metastable distances between freely movable plates immersed into a fluid reservoir (Fig. 4). Furthermore, within the Derjaguin approximation $\gamma(L)$ renders the force between two large spheres suspended in a liquid of small spheres (see Eq. (4.14)).

(6) Within the LWDA and based on the above results we have determined the direct correlation function. By inverting the Ornstein-Zernicke equation the total correlation function has been calculated. We have discussed it within the framework of the Percus test-particle theorem (Fig. 6).

(7) The influence of the distant wall on the radial distribution function $g(r_{12}, z_1 = \sigma, z_2 = \sigma)$ along the near wall decreases rapidly with increasing slit width $L$ (see Fig. 7).

(8) For a fixed slit width and bulk reference density $\rho_b$ we have analyzed the dependence of the radial distribution function $g(r_{12}, z_1, z_2 = z_1)$ on $z_1$. We find that for those values of $z_1$ for which the density profile $\rho(z)$ exhibits a local minimum (maximum) the amplitude of the oscillations of this correlation function as function of $r_{12}$ is enhanced (reduced) compared to the corresponding bulk correlation function.

A major advantage of the present density functional theory is that it is computationally much less demanding than integral equation theories or numerical simulations without loosing its competitiveness as far as the quantitative reliability is concerned. Furthermore density-functional theory yields relatively easy access to free energies and solvation forces. Therefore we are encouraged to extend this analysis to fluids governed by dispersion forces in order to refine the presently available ansatz for the two-point correlation functions in such systems [47].

## ACKNOWLEDGMENTS

We thank Professor H. Löwen and Dr. M. Schmidt for useful discussions regarding Sec. IV.C.

## APPENDIX:

Upon differentiating the equilibrium grand canonical potential in Eq. (2.6) one obtains by using the chain rule

$$\frac{\delta \Omega}{\delta V(\mathbf{R})} = \rho(\mathbf{R}) + \int_{\mathbb{R}^d} d^d R \left\{ \frac{\delta F[\rho]}{\delta \rho(\mathbf{R}')} - (\mu - V(\mathbf{R}')) \right\} \frac{\delta \rho(\mathbf{R}')}{\delta V(\mathbf{R})}. \tag{A1}$$

For the equilibrium density distribution the expression within the curly bracket vanishes. This is true even for *approximate* expressions for the functional $F[\rho] = F_{id}[\rho] + F_{ex}[\rho]$ like, e.g., the one used for the LWDA. According to Eq. (2.10) the external potential $V(\mathbf{R})$ depends parametically on $L$ so that with Eq. (A1) one has

$$\begin{aligned}\left(\frac{\partial \Omega}{\partial L}\right)_{T,\mu,A} &= \int_{\mathbb{R}^d} d^d R \frac{\delta \Omega}{\delta V(\mathbf{R})} \frac{\partial V(\mathbf{R})}{\partial L} \\ &= \int_{\mathbb{R}^d} d^d R \, \rho(\mathbf{R}) \, (-\frac{1}{\beta} \exp(\beta V(\mathbf{R})) \frac{\partial \exp(-\beta V(\mathbf{R}))}{\partial L}) \\ &= -\frac{1}{\beta} \int_{\mathbb{R}^d} d^d R \, \rho(\mathbf{R}) \exp(\beta V(\mathbf{R})) \frac{\partial}{\partial L} (\Theta(z - \sigma) + \Theta(L - \sigma - z) - 1)\end{aligned}$$

$$\begin{aligned}
&= -A\frac{1}{\beta}\left(\rho(z)\exp(\beta V(z))\right)_{z=L-\sigma} \\
&= -A\frac{1}{\beta}\rho(L-\sigma) = -A\frac{1}{\beta}\rho(\sigma).
\end{aligned} \tag{A2}$$

The last but one step in Eq. (A2) is based on the fact that the product $\rho(z)\exp(\beta V(z))$ is continuous as function of $z$ [49] so that the value of this product at $z = L-\sigma$ can be obtained by considering the limit $z \to L-\sigma-0$ where $V(z) = 0$. This renders the equivalence between the two definitions in Eqs. (4.8) and (4.12) both for exact and for approximate expressions for $F[\rho]$.

# FIGURES

FIG. 1. Density profile of a fluid of hard spheres with diameter $\sigma$ between two hard walls located at $z = 0$ and $z = L$ (compare Eq. (2.10)) according to the LWDA density functional theory (full curve) for the slit widths $L = 5.1\sigma$ (i.e. $\tilde{L} = 3.1\sigma$) (a) and $L = 3.8\sigma$ (i.e. $\tilde{L} = 1.8\sigma$) (b). In (a) the corresponding bulk density is chosen as $\rho_b \sigma^3 = 0.546$ to allow comparison with molecular dynamics simulation data [13]. In (b) the bulk density is $\rho_b \sigma^3 = 0.683$ as in all remaining figures.

FIG. 2. Within the LWDA the contact density $\rho(\sigma)$ (full line) and the density $\rho(L/2)$ in the center of the slit (dotted line) of a hard sphere fluid are shown for a bulk density of $\rho_b \sigma^3 = 0.683$. The slit width $L$ varies between $2.1\sigma$ and $8\sigma$. In the limit $L \to \infty$ the contact density approaches the constant value $\rho_\infty(\sigma)\sigma^3 = 3.84$. In the limit $L \to 2\sigma$ the two densities $\rho(\sigma)$ and $\rho(L/2)$ approach the same limit $\Lambda^{-3} \exp(\mu/kT)$ as given by Eq. (3.3b). The dots denote results of grand canonical Monte Carlo simulations [11].

FIG. 3. The coverage $\Gamma(L)$ and the finite size contribution to the free energy $\gamma(L)$ (Eq. (4.6)) of a hard sphere fluid between two flat hard walls for various slit widths $L = 0.1\sigma \ldots 8\sigma$ at a bulk density of $\rho_b \sigma^3 = 0.683$. In the limit $L \to \infty$ the coverage approaches $\Gamma(\infty) = -0.92\sigma^2$ and $\gamma(L)$ reaches the value $\beta\gamma(\infty)\sigma^2 = -1.95$.

FIG. 4. Schematic plot of a model system consisting of a planar hard wall ($a$) on the left side and a hard piston ($b$) which is in front of and parallel to the left wall. It can be fixed at different distances $L$. The hard sphere fluid between $a$ and $b$ and to the right of $b$ are connected to the same grand canonical reservoir and thus are in equilibrium with each another. The force per area and $k_B T$ acting on the piston is given by $\beta f(L) = \rho(\sigma) - \rho_\infty(\sigma)$ and is plotted in Fig. 2 as the full line.

FIG. 5. The force $f_s(h)$ between two fixed hard spheres of a large radius $R$ immersed into a fluid of hard spheres of diameter $\sigma \ll R$ as function of the minimum distance $h$ between the surfaces of the large spheres (see Eq. (4.14)). The centers of the large spheres are at a distance $2R + h - \sigma$. The dotted curve represents the effective interaction potential between the two large spheres.

FIG. 6. Within LWDA the conditional singlet density $\rho(\mathbf{R}_1|\mathbf{R}_2) = \rho(z_1) \, g(r_{12}, z_1, z_2 = \sigma)$ of a hard sphere fluid in a slit of width $L = 5.1\sigma$ at a reference bulk density of $\rho_b \sigma^3 = 0.546$ in the presence of a hard sphere of the same diameter whose center is fixed at $\mathbf{R}_2 = (r_{12} = 0, z_2 = \sigma)$. (This position is marked by a cross.) Due to $g(r_{12} \to \infty, z_1, z_2) = 1$ one recovers for $r_{12} \to \infty$ Fig. 1(a) which corresponds to the same values of $L$ and $\rho_b$. The dots and circles in the contour line plot at the bottom of the figure denote the positions of the local maxima and minima, respectively. The dashed line indicates the excluded volume due to the test particle; it is not an isodensity line. The value of the singlet density varies by an amount of $0.1\sigma^{-3}$ between neighboring contour lines. The contour lines are shown only for values less then $1.0\sigma^{-3}$. (In the actual calculations a mesh size of $0.02\sigma$ in the $z$- and the $r_{12}$-direction has been used.) As can be seen from the contour lines the perturbation of the density distribution in the slit due to the presence of the fixed sphere at $\mathbf{R}_2 = (0, \sigma)$ dies out for $r_{12} \gtrsim 3\sigma$ or $z_1 \gtrsim 3\sigma$.

FIG. 7. The radial distribution function $g(r_{12}, z_1, z_2)$ of a hard sphere fluid for various slit widths $L$ at the reference bulk density $\rho_b \sigma^3 = 0.683$. The lateral distance $r_{12}$ varies between $1.2\sigma$ and $3.0\sigma$ for $z_1 = z_2$ fixed with $z_1 = \sigma$ in (a) and $z_1 = 1.5\sigma$ in (b). In the limit $r_{12} \to \sigma$ the correlation functions increase rapidly to values $2.3 \ldots 2.4$ in (a) and $2.8 \ldots 3.1$ in (b) depending on $L$. Note that we have used the same scales of the axes in (a) and (b) in order to facilitate a direct comparison of the two cases. In both plots the dashed–double-dotted curves denote the correlation function of the corresponding homogeneous bulk fluid within the PY-approximation. In (a) for all values of $r_{12}$ the correlation function cannot be distinguished from its semi-infinite form [22] for $L \gtrsim 5\sigma$.

FIG. 8. The radial distribution function $g(r_{12}, z_1, z_2 = z_1)$ of a hard sphere fluid in a slit of width $L = 5.1\sigma$ at a bulk density $\rho_b \sigma^3 = 0.546$. The decay of the correlations parallel to the walls is shown for various values of $z_1 = z_2$. According to Fig. 1(a) $z_1 = 1.0\sigma$ corresponds to the contact with the left wall, $z_1 = 1.64\sigma$ to the first minimum, $z_1 = 2.2\sigma$ to the first maximum, and $z_1 = 2.55\sigma$ to the midpoint of the density profile of this system.

FIG. 9. The radial distribution function $g(r_{12}, z_1, z_2 = z_1)$ of a hard sphere fluid in a slit of width $L = 3.8\sigma$ at a bulk density $\rho_b \sigma^3 = 0.683$. As in Fig. 8 the decay of the correlations parallel to the walls, i.e. as function of $r_{12}$, is shown for various values of $z_1 = z_2$. According to Fig. 1(b) $z_1 = 1.0\sigma$ corresponds to the contact with the left wall, $z_1 = 1.46\sigma$ to the first minimum, $z_1 = 1.78\sigma$ to the first maximum, and $z_1 = 1.9\sigma$ to the midpoint of the density profile of this system.

TABLES

TABLE I. Characteristics of a hard sphere fluid between two hard walls of width $L$ at a bulk density $\rho_b \sigma^3 = 0.683$. The table lists those slit widths for which the contact density $\rho(\sigma)$, the density $\rho(L/2)$ at the center, the coverage $\Gamma(L)$ (Eq. (4.1)), and the finite size contribution to the free energy $\gamma(L)$ (Eq. (4.6)) attain their extrema.

|  | width $L/\sigma$ leading to minima of | | | | | |
| --- | --- | --- | --- | --- | --- | --- |
| $\rho(\sigma)$ | 2.39 | 3.41 | 4.42 | 5.43 | 6.43 | 7.42 |
| $\rho(L/2)$ |  | 3.07 |  | 5.12 |  | 7.12 |
| $\Gamma(L)$ | 2.63 | 3.62 | 4.60 | 5.60 | 6.60 | 7.60 |
| $\gamma(L)$ | 2.18 | 3.12 | 4.22 | 5.21 | 6.21 | 7.20 |
|  | width $L/\sigma$ leading to maxima of | | | | | |
| $\rho(\sigma)$ |  | 3.01 | 3.98 | 4.95 | 5.95 | 6.93 |
| $\rho(L/2)$ |  |  | 4.12 |  | 6.15 |  |
| $\Gamma(L)$ | 2.06 | 3.10 | 4.10 | 5.10 | 6.09 | 7.08 |
| $\gamma(L)$ | 2.76 | 3.74 | 4.72 | 5.71 | 6.71 |  |

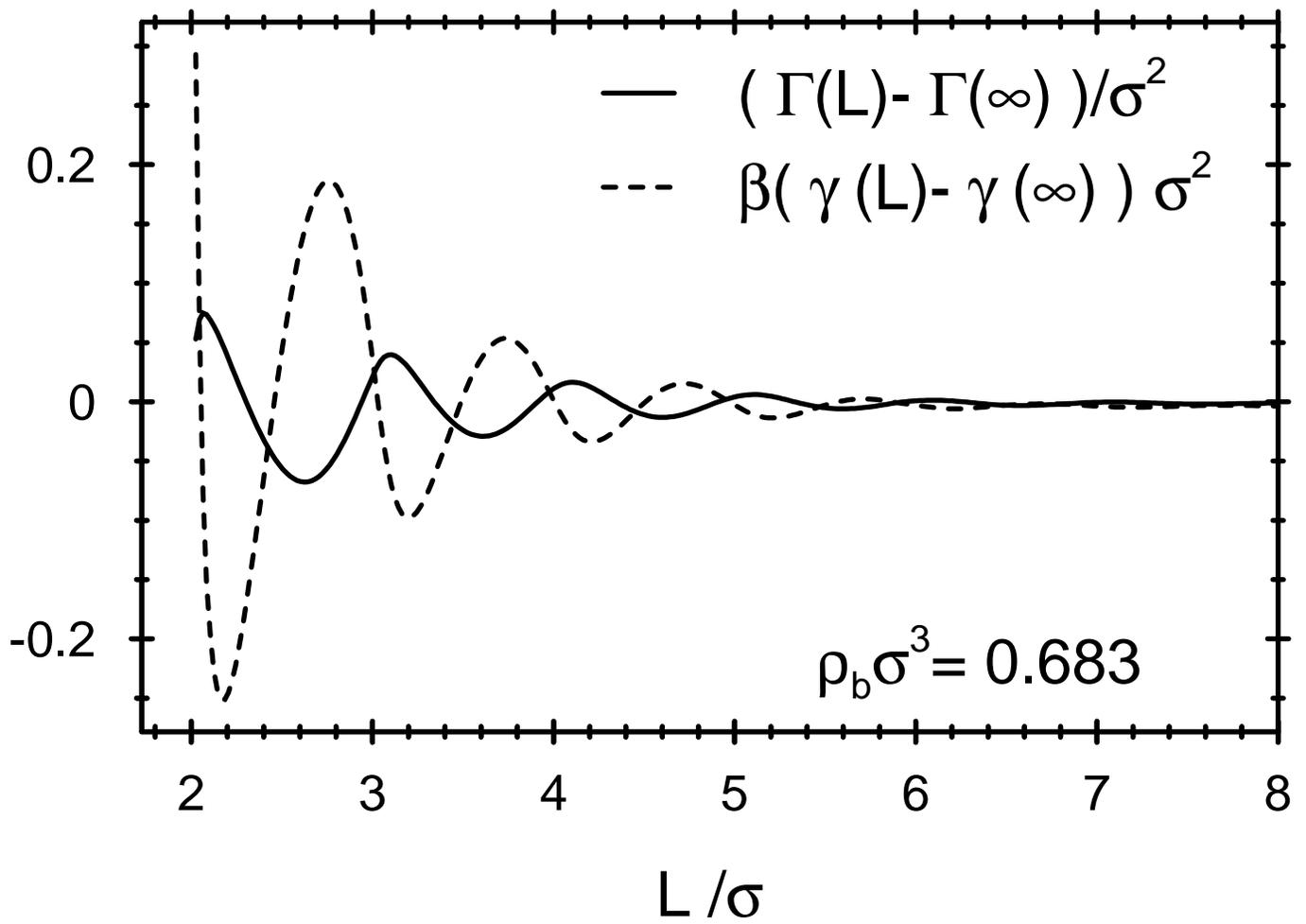

*Fig. 3*

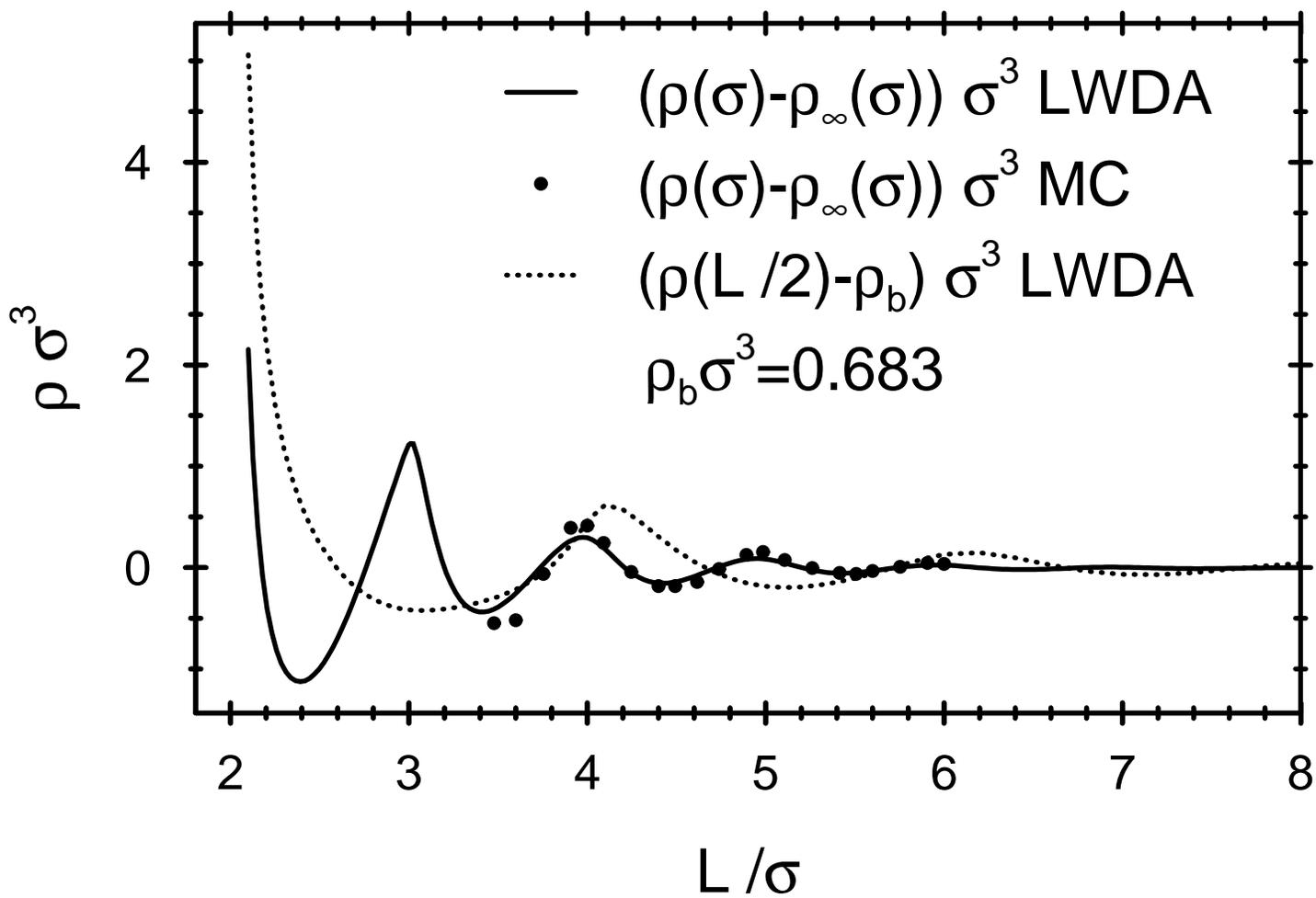

*Fig. 2*

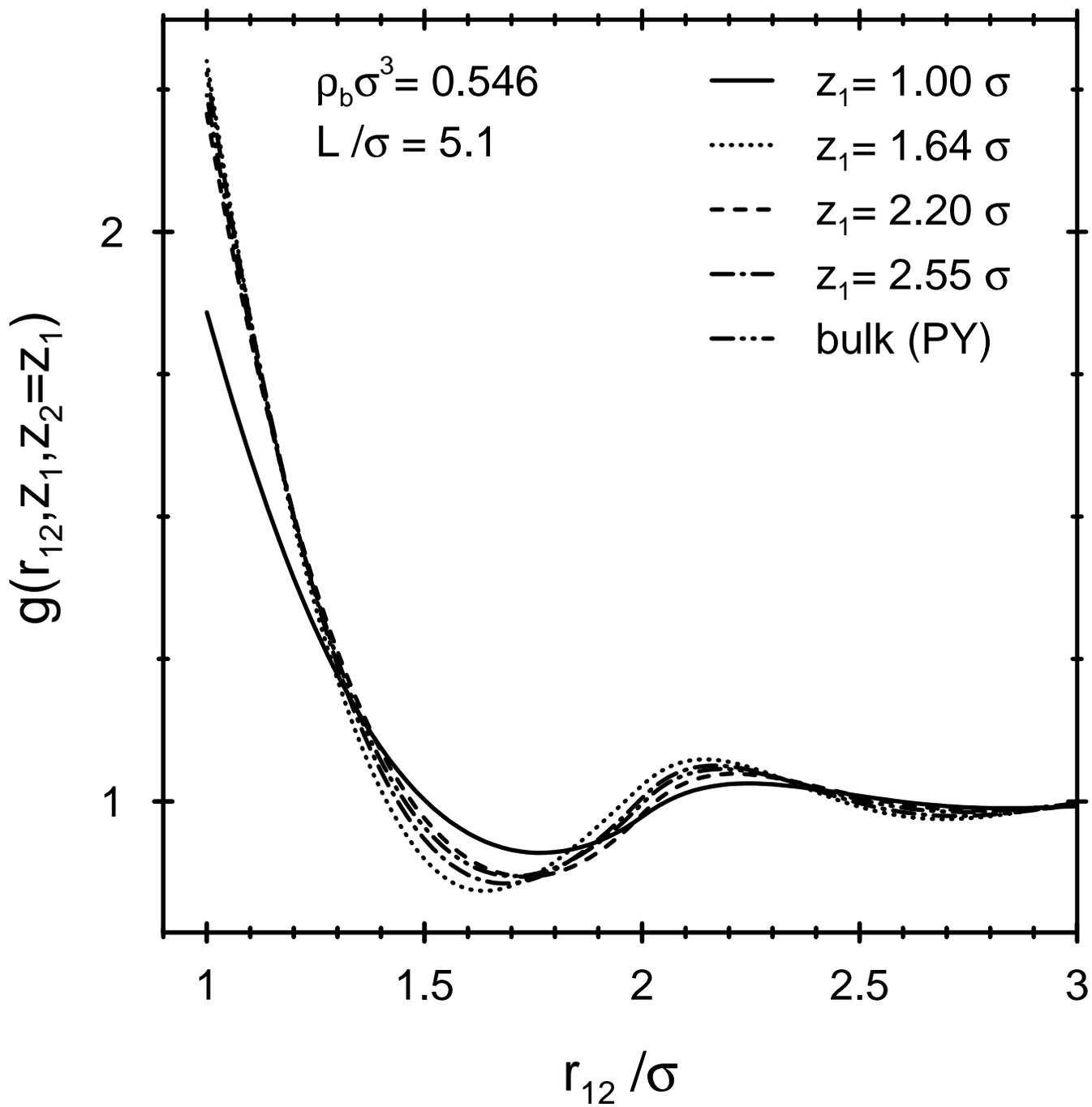

*Fig. 8*

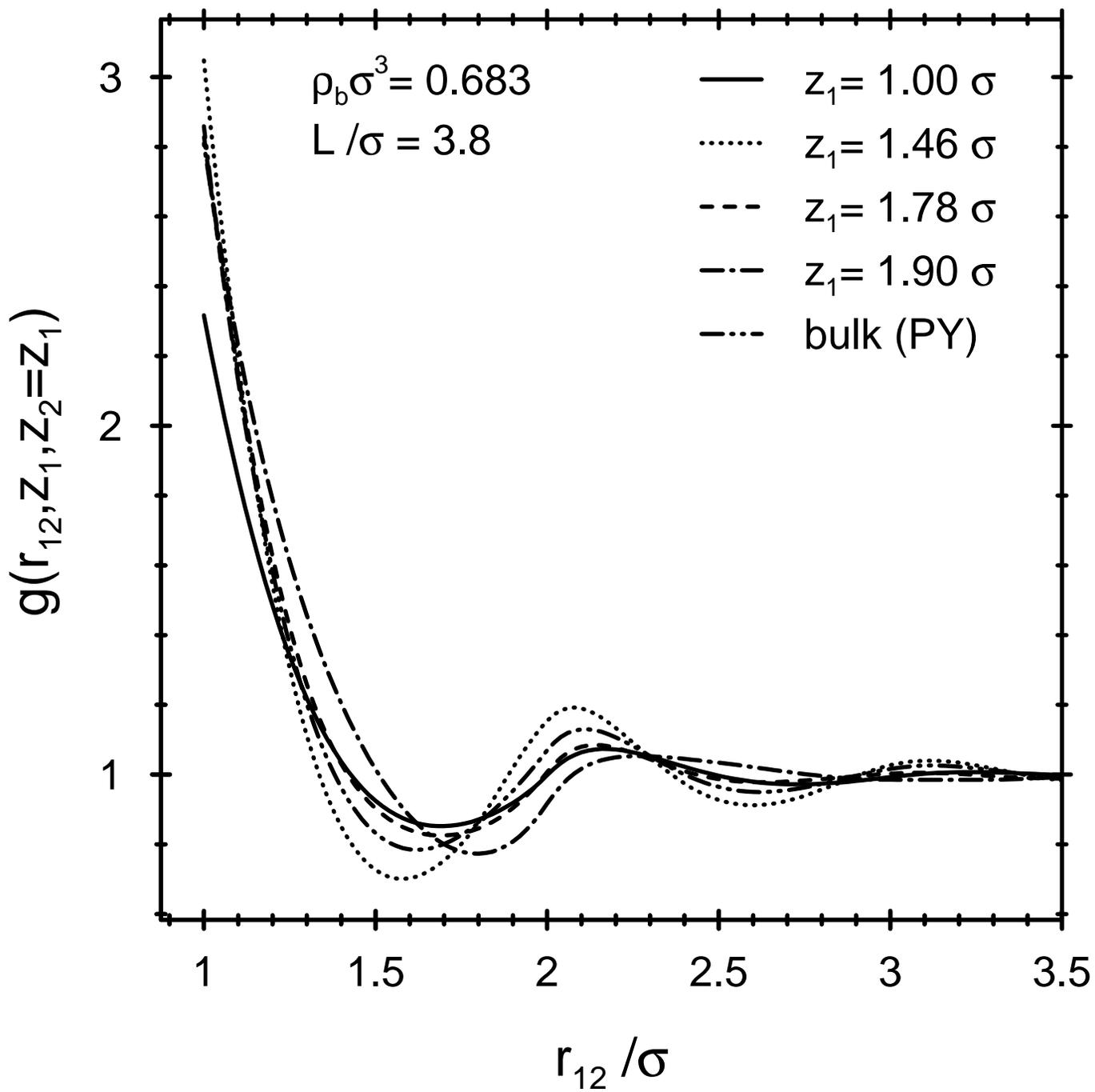

*Fig. 9*

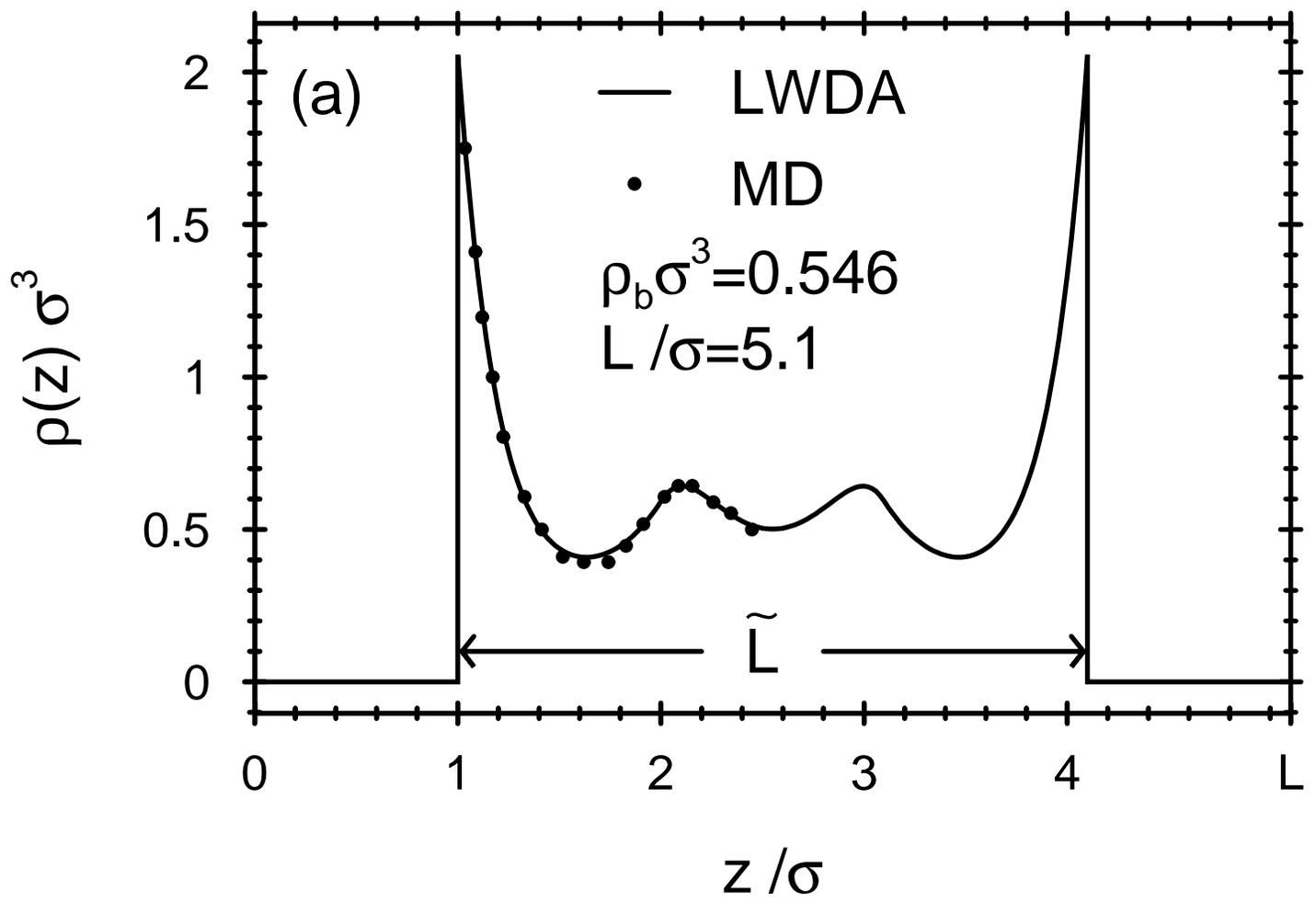

*Fig. 1a*

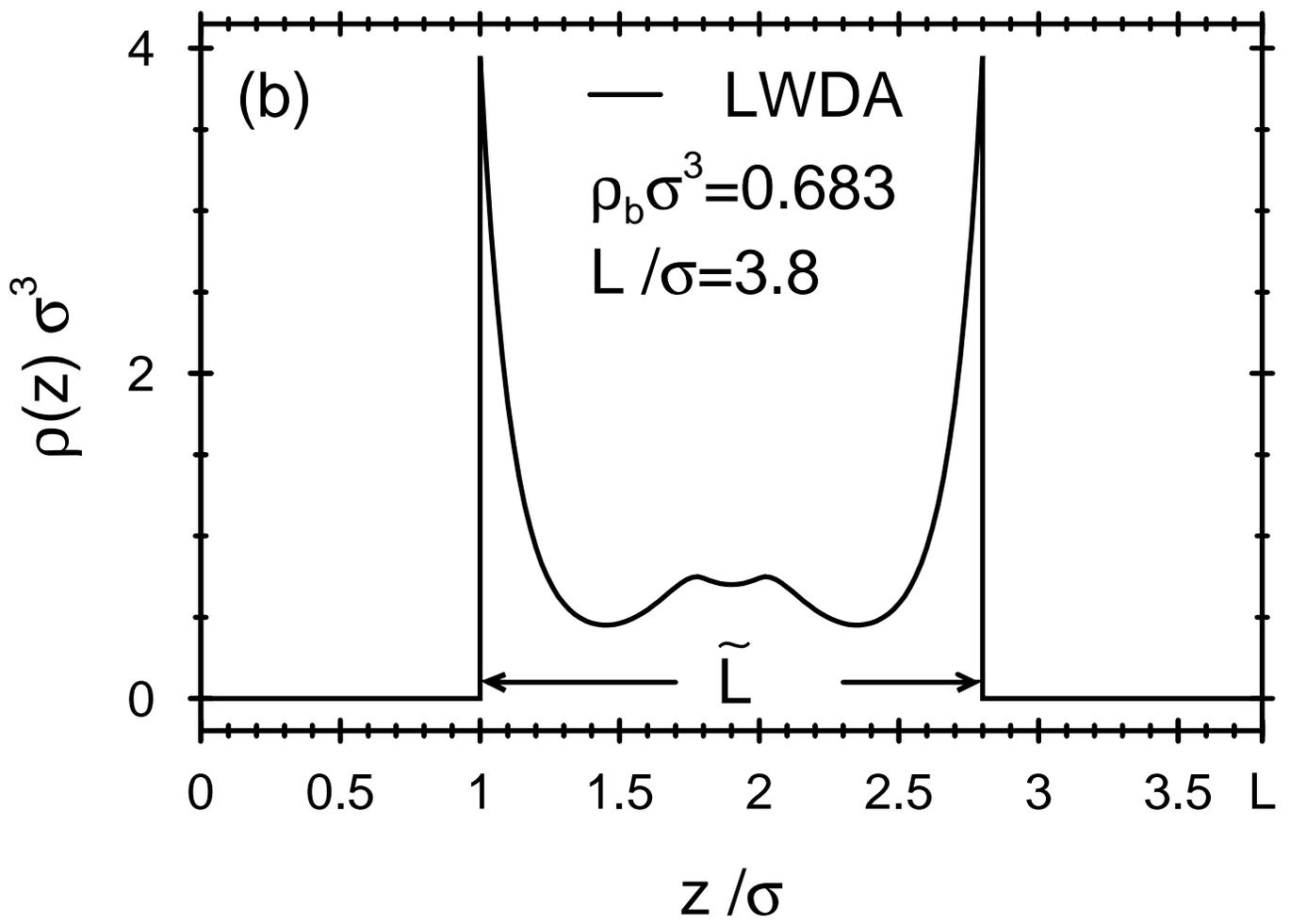

*Fig. 1b*

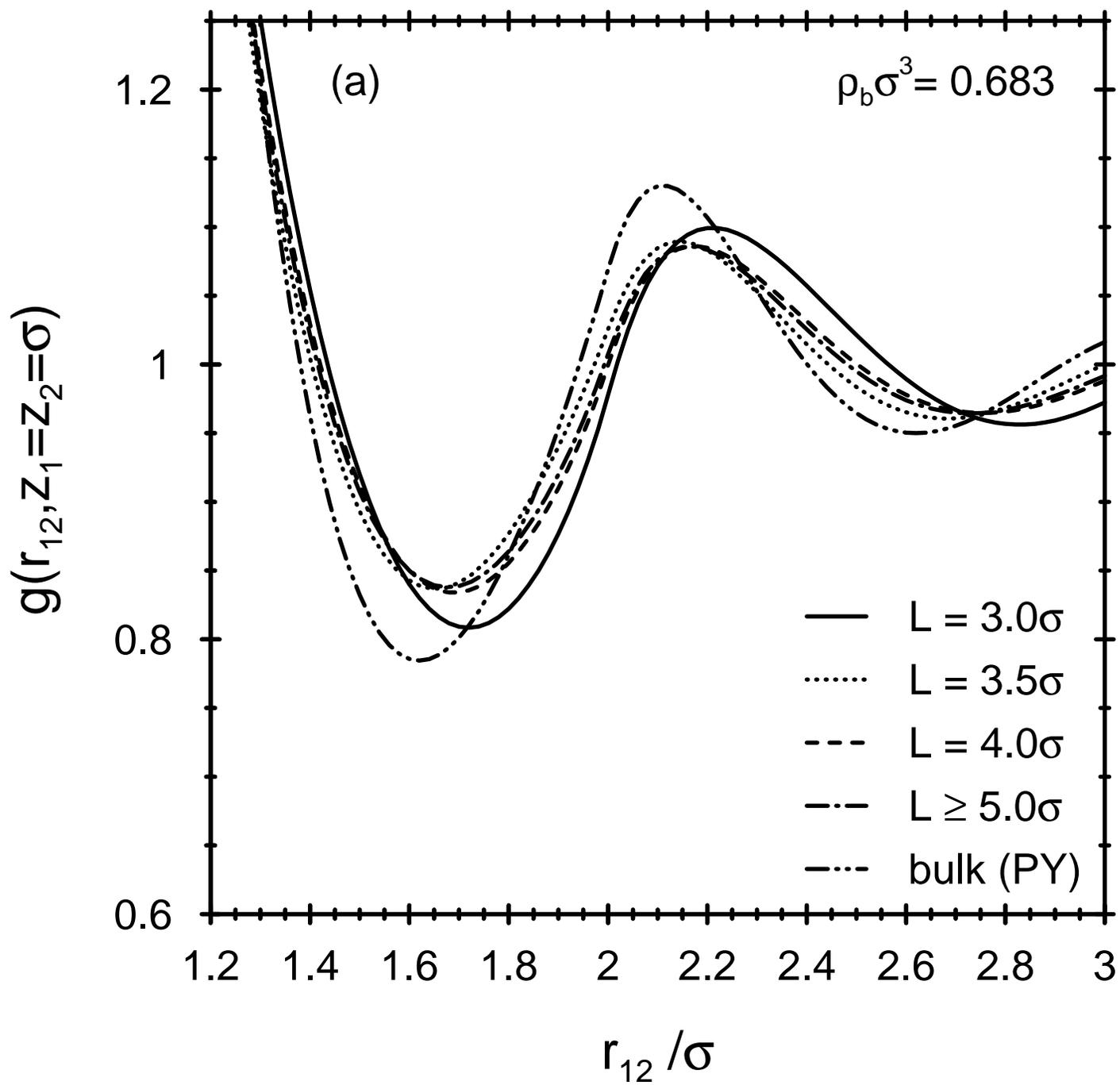

*Fig. 7a*

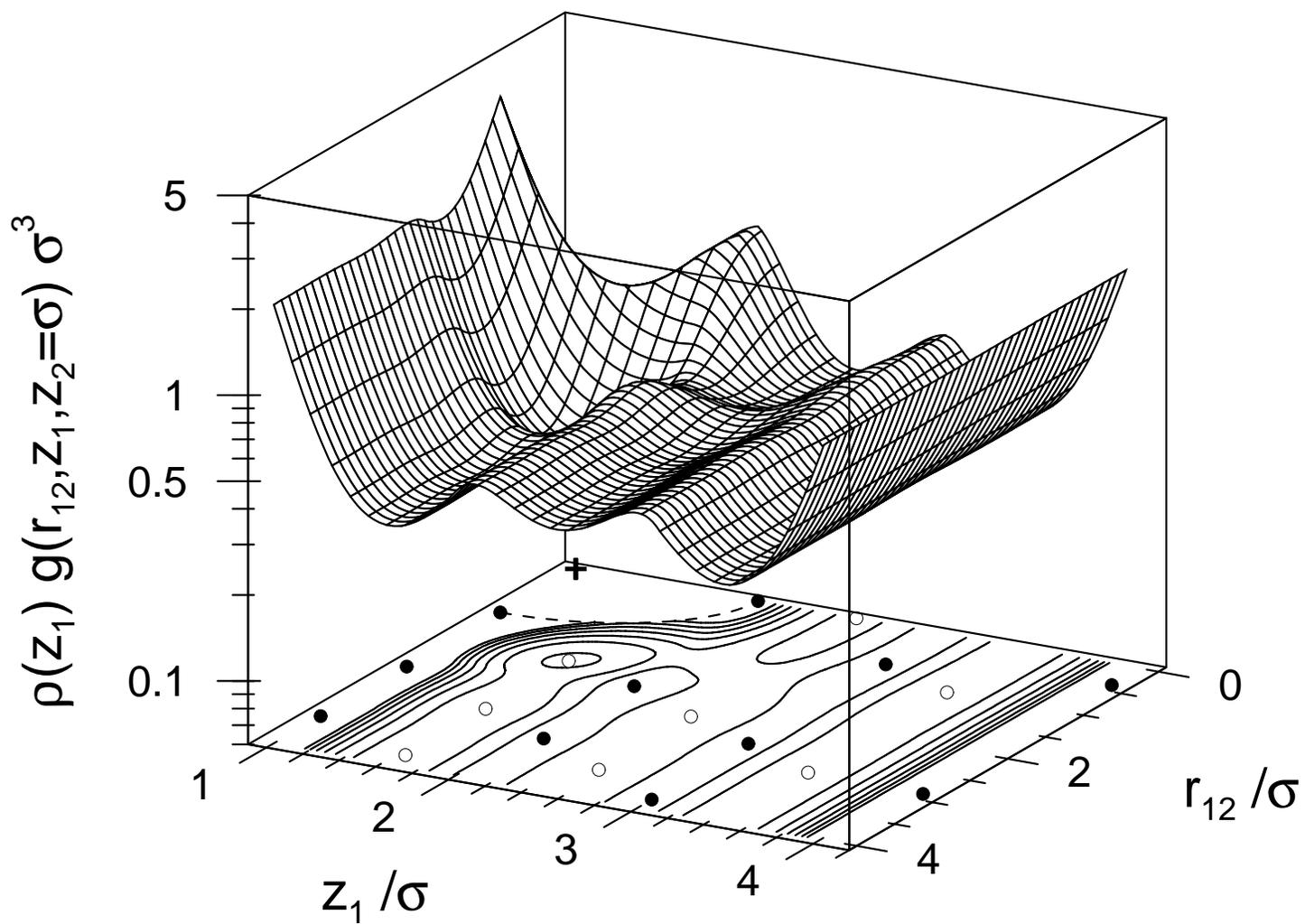

Fig. 6

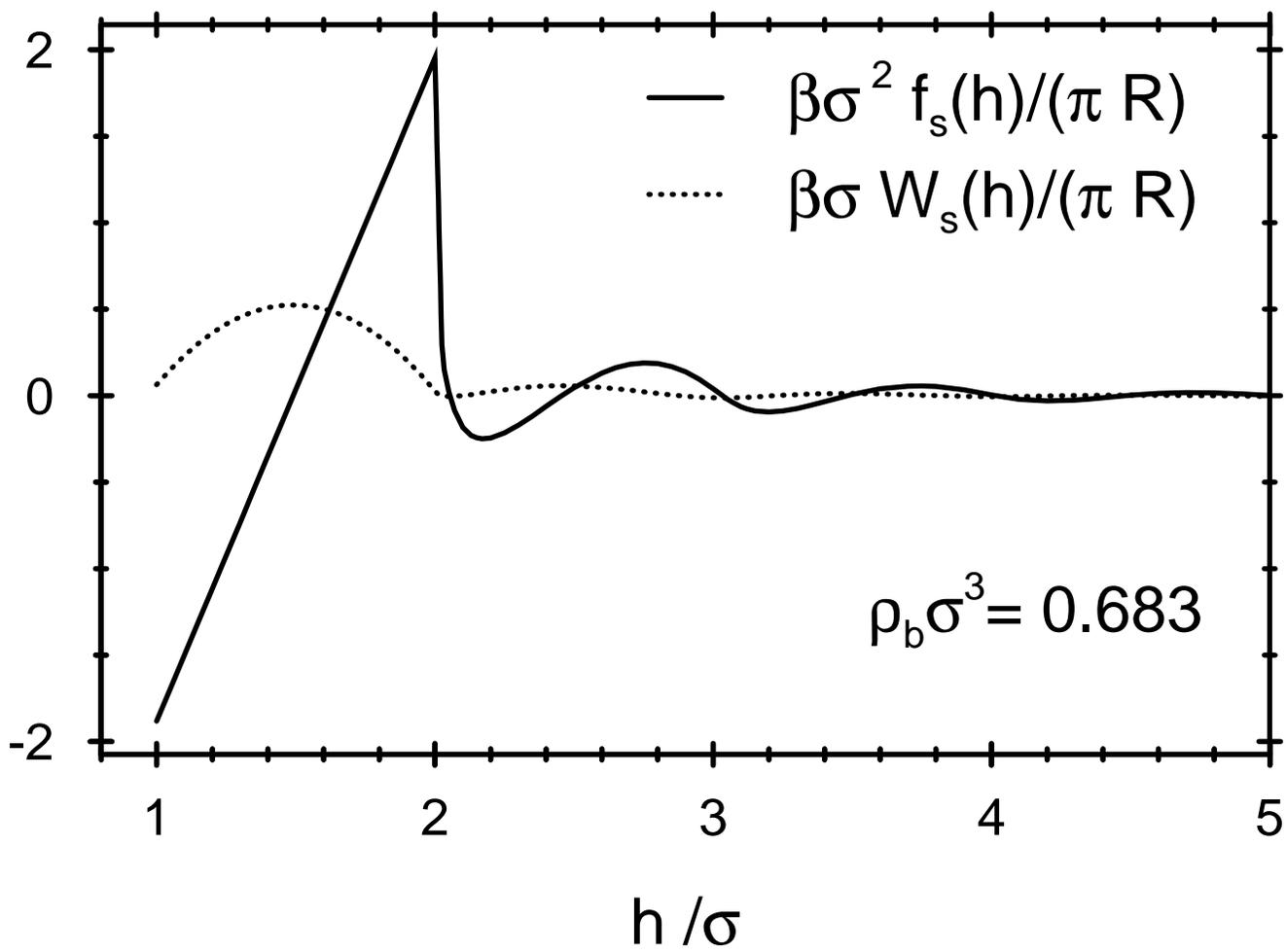

*Fig. 5*

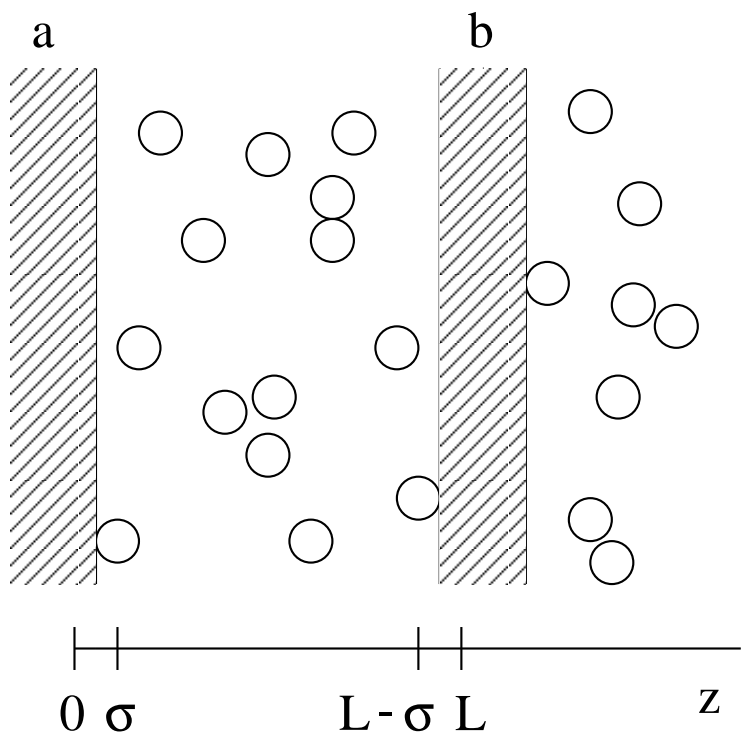

Fig. 4

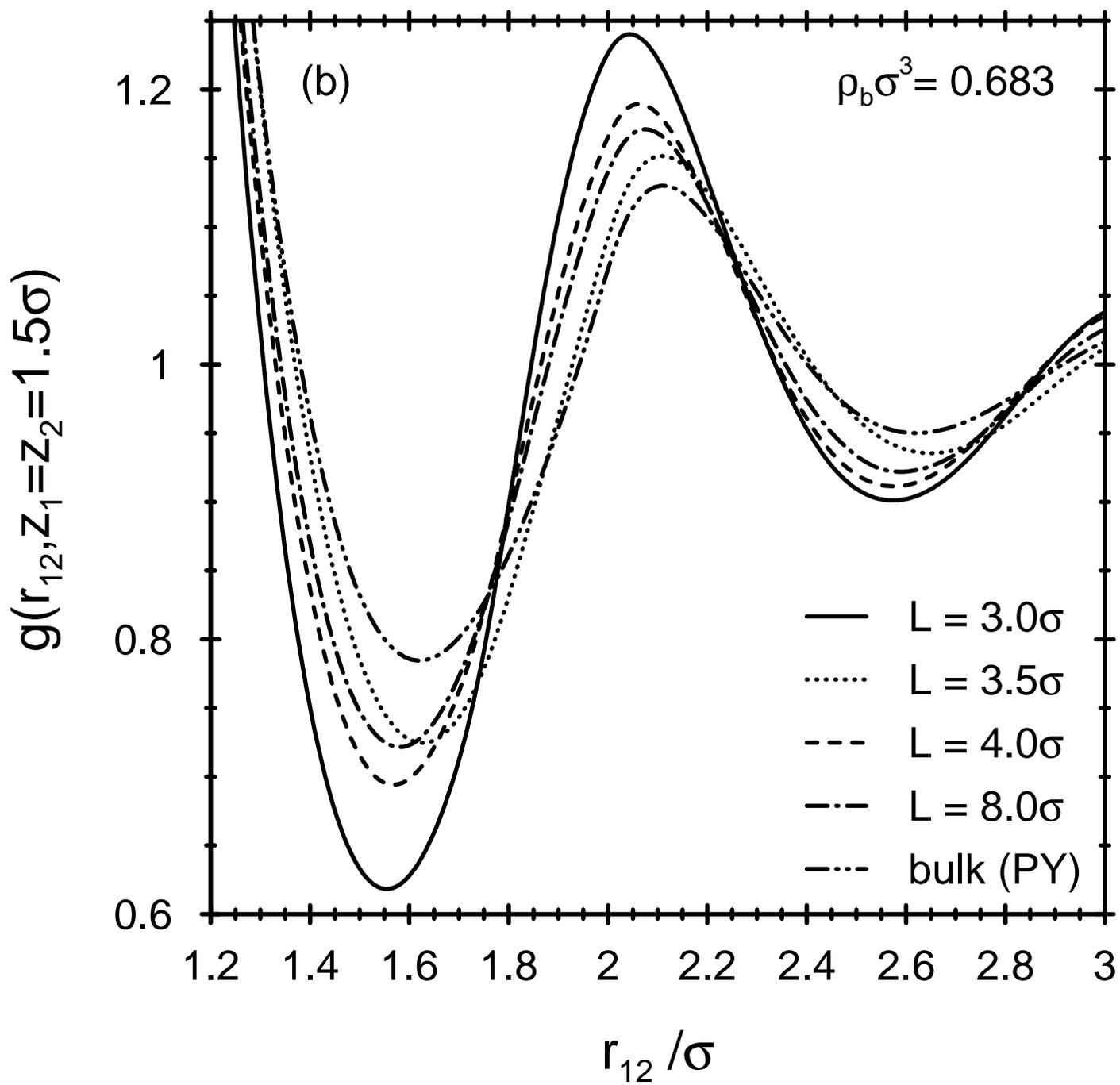

*Fig. 7b*